\documentstyle[aps]{revtex}
\begin{document}
\twocolumn[
\draft
\tighten
\preprint{HU-SEFT R 1994-15}
\title{Proton-ring and Electron-linac Collider (PRELC)\\
as a (first) TeV-range\\
electron-proton or photon-proton collider}
\author{\bf Raimo Vuopionper\"a}
\address{Research Institute for High Energy Physics,
University of Helsinki,\\
P.O.Box 9, FIN-00014 University of Helsinki, FINLAND\\
E-mail: vuopionp@pcu.helsinki.fi (Internet)\\ ${}^{}$}
\maketitle
\widetext
\begin{abstract}
The use of the existing proton storage rings combined with  electron linear
accelerator as a ring-linac type electron-proton or photon-proton collider is
investigated. The total CM-energy of {\em Proton-ring and Electron-linac
Collider} (PRELC) is in the range of $1.0$ TeV. The most important physical
issues are listed and the most critical machine aspects of the PRELC are
studied. It is shown that the luminosities in the range of $10^{31}$ to
$10^{32}$ could be achieved. The PRELC could be used simultaneously with the
$e^+e^-$-collider if the refocused electron beam could be used as the electron
beam for the PRELC.
\end{abstract}
]
\narrowtext
The verification of the quark-parton model has been one of the successes of
the $e^-p^+$-collisions. The energy limitations of the fixed-target
experiments and the results of HERA \cite{HERA} electron-proton collider have
generally increased the interest to the electron-proton colliders. Due to the
energy losses caused by synchrotron radiation at electron rings the future of
storage ring type electron-proton colliders seems to be unrealistic and
therefore one must consider ring-linac type colliders
\cite{old,otherep,otheree,othergp,gammaphys,RKJPRV} as an option. Furthermore,
the PRELC ({\em Proton-ring and electron-linac Collider}) also offers us the
option to have TeV range photon-proton collisions \cite{othergp,gammaphys}.
The high energy photon beam is produced by the Compton backscattering of laser
photons off the high energy electron beam.

\begin{figure}[h]
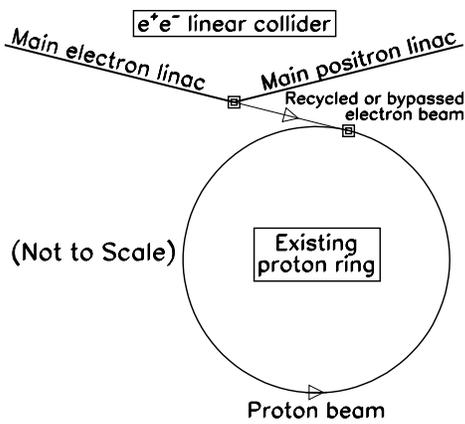

\caption{\rm Schematic view of proton-ring and electron-linac collider.}
\label{fig:kuva}
\end{figure}

In this paper the possibility to construct electron-proton or photon-proton
collider using the existing proton collider rings with linear electron
accelerator of the proposed $e^-e^+$ linear colliders is investigated.
Studying the most critical machine aspects for this kind of ring-linac
colliders we have been able to outline very robust and general
principles for a realistic design proposal for the PRELC.
Four different types of PRELC are presented, where the existing
proton storage ring or the extracted proton beam from this ring is
combined with the electron beam the $e^-e^+$-collider or
combined with the photon beam (Compton backscatterred laser beam).
The possible options for the PRELC are shown in Table \ref{T:pos_option}
and the most crucial machine parameters for two options are listed in Table
\ref{T:posi_param}. The schematic view of PRELC is presented in Fig.
\ref{fig:kuva}. We also outline the most important physical issues which
can be studied at the PRELC (see Tables \ref{T:phys_ep} and \ref{T:phys_gp}).

\begin{figure}[h]
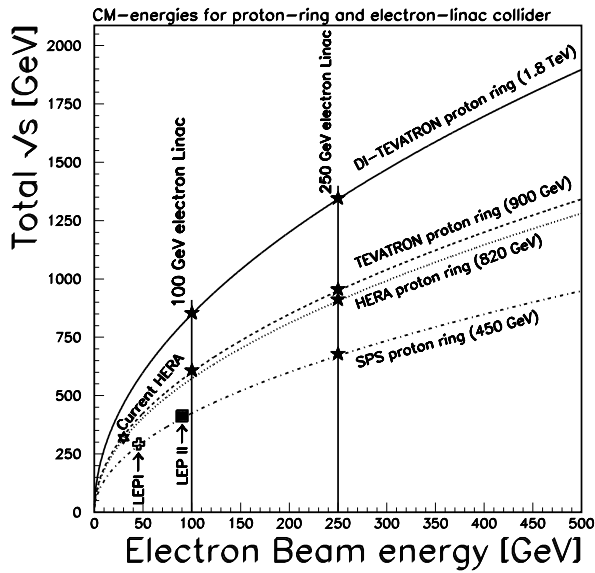

\caption{\rm Possible values of total $\protect\sqrt{s}$ of the proton-ring and
electron-linac collider.}
\label{fig:tots}
\end{figure}

 From the machine point of view one has four different collider options at
the PRELC: two different machine layouts for electrons and photons.
One could collide the circulating proton beams with beams from the linac or
one could arrange collisions between the extracted proton beams and beams
from the linac. The first option is limited by the proton ring lattice and
the beam characteristics of the proton ring (e.g. proton beam emittance,
properties of proton ring lattice, etc.). For both options the luminosity of
the collider is:
\begin{equation}
{\cal L} =
\frac{N_pN_e f}{4\pi \sigma_x \sigma_y} \times \epsilon,
\end{equation}
where $\sigma_x=\max (\sigma^e_x,\sigma^p_x)$,
$\sigma_y=\max (\sigma^e_y,\sigma^p_y)$, $f= \min (f_e,f_p)$
and $\epsilon$ is the 'efficiency' of the collider. For simplicity we
assume $\epsilon = 1.0$ except when the proton and electron (photon) beam
cross-sections are very unequal in size reflecting the higher particle
densities (gaussian) in the central regions of the particle bunches,
in which case we take $\epsilon = 2.0$. The possible options for the PRELC
are given in Table \ref{T:pos_option} (see also Fig. \ref{fig:tots}).

In the case of the
circulating proton beam the luminosity is limited by the proton beam
characteristics, namely, by the proton beam spotsize (see Table
\ref{T:posi_param}). Furthermore, the optimal frequency matching is not very
easy at circulating proton beam option since the particle bunch
intervals of the proton rings and electron linacs do not match (see Table
\ref{T:posi_param}). Using the 'design' values for both proton and electron
beams (given in the Refs. \cite{PPARA,EPARA}) the maximun luminosity would be
between $1\ldots 3 \times 10^{28}\;
{\rm s^{-1} cm^{-1}}$ and $4\ldots 8 \times 10^{29}\; {\rm s^{-1} cm^{-1}}$
for different PRELC 'proposals'.

Of course one can change some of the proton beam characteristics, like the
beam emittances, by building new modern injectors to the existing storage rings
or by rebuilding (completely or partially) the proton ring lattice. The problem
is that these changes could be quite expensive and furthermore, in the first
case it might not be enough to lower the emittance of the injected beam
since the characteristical values of the beams at the storage rings have
the tendency to move towards the
ones determined by the ring lattice.

We can study the effects of these small modifications
to the proton beam and the proton ring lattice
using the formula for the beam size $
\sigma_{x(y)} = \sqrt{\varepsilon_{x(y)} \times \beta_{0,\, x(y)}} \:$,
where $\varepsilon$ is the emittance and the $\beta_0$
is the focal point value of the magnetic $\beta$-function, at the
interaction point.
If we can change the normalized emittance
$\varepsilon_{N} = \frac{p}{m_0c} \varepsilon
$ of the proton beam to
$\varepsilon_{N} =
1.0\times 10^{-6}$ $\pi$ rad-m, design value of SSC  \cite{PPARA},
the luminosity of the PRELC will increase by a factor of $5\ldots 9$,
depending on the design option. The maximum luminosity would be between
$1\ldots3 \times 10^{29}\; {\rm s^{-1} cm^{-1}}$ and $2\ldots 4
\times 10^{30}\; {\rm s^{-1} cm^{-1}}$. One could also change the
value of the magnetic $\beta$-function at the interaction point and one should
realistically be able to gain additional factor of $2\ldots 5$ which would
increase the luminosity of the PRELC maximally to the level of $ 10^{31}
\; {\rm s^{-1} cm^{-1}}$.

In the case of the extracted proton beam the proton ring
lattice limitations do not play a big role, since the proton beam need not be
stable after the collision. This option has also the advantage that we would
be able to adjust the collision frequency by extracting the proton
bunches from
the proton ring at the
appropriate time and the optimal ring position. Ideally, one
would be able to run the proton-collider simultaneously.

Now using the improved
invariant emittance $\varepsilon=1.0\times 10^{-6}$ $\pi$ rad-m for the
proton beam, stronger focusing (i.e. smaller $\beta_0$) at the
collision point and other methods (e.g. increasing the proton beam current) we
can improve the luminosity of the collider realistically
to the level of $10^{31}\; {\rm s^{-1} cm^{-1}}$ (optimitically to ${\cal L}
\approx 10^{32}\; {\rm s^{-1} cm^{-1}}$).

For example using
values $\varepsilon=1.0\times 10^{-6}$ $\pi$ rad-m and $\beta_0=0.1$ m at the
TEVATRON-TESLA option the luminosity of the PRELC would be ${\cal L}
\approx 1.0\times 10^{31}\; {\rm s^{-1} cm^{-1}}$. In order to reach the
luminosities of the order of $10^{32}\; {\rm s^{-1} cm^{-1}}$ one must
really try to 'fine-tune'
the proton beam
characteristic values (e.g. the proton bunches must be made shorter, more
intense etc.), which seems to be quite expensive and difficult.

All the results presented above apply to the electron-proton collider options
as well as to the photon-proton collider option. In the photon-proton collider
option the photon beam is obtained by Compton backscattering of laser beam off
the high energy electron beam. The electron beam conversion to photon beam is
very efficient ($\approx 100$\%), i.e. the conversion coefficient is $1.0$,
which means that $N_\gamma = N_{e^-}$.

One very interesting possibility is to use a spent and refocused beam
(recirculated beam) of the $e^+e^-$-collider as an electron beam for the PRELC.
This recirculation scheme
is possible for the TESLA $e^+e^-$
linear collider \cite{TESLARE}.
Actually, it does not matter if the characteristic values
(e.g., emittance, etc.) of the electron beam are worse after the refocusing
process since the spotsize is determined by the proton beam spotsize at the
collision point.
This option makes possible to use the PRELC and $e^+e^-$-collider
simultaneously, especially, for the photon-proton collider
this
seems to be very interesting option.

A typical reaction in the electron-proton collider is
$e^- p^+ \rightarrow \ell + X $, where $\ell$ can be either charged lepton or
neutrino and $X$ is some specified final state. The kinematics of such event
is described by following equations:
\begin{equation}
\begin{array}{lclcl}
Q^2 &=&-q^2 &\approx & 4E_eE_{\ell}\sin^2(\frac{\theta_{\ell}}{2})
\;,\vspace{0.1 cm}\\
\gamma_{{}_{CM}} &= &\frac{E_p + E_e}{\sqrt{s}} &= &
\frac{1}{2} \left(\sqrt{\frac{E_p}{E_e}} +
\sqrt{\frac{E_e}{E_p}}\, \right)\;,
\vspace{0.1 cm}\\
m_p\nu &=&p_p\cdot q &\approx & 2E_p\left[E_e - E_{\ell} \cos^2
(\frac{\theta_{\ell}}{2}) \right]
\;,\vspace{0.1 cm}\\
x &= &\frac{Q^2}{2p_p\cdot q} &\approx &
\frac{E_eE_{\ell}\sin^2(\frac{\theta_{\ell}}{2})}{E_p[E_e-
E_{\ell} \cos^2 (\frac{\theta_{\ell}}{2}) ]}\;,
\end{array}
\label{invar1}
\end{equation}
where $s \approx 4E_eE_p$ and $\theta_{\ell}$ is the lepton scattering angle.
The Bj{\o}rken variable $x$ obeys $0 \geq x \geq 1$.
 If $E_p > E_e$, then
the CM frame is moving to the direction of proton beam,
and in the opposite case to the direction of electron beam.
Using the fact that in CM-frame
$Q^2
\approx (s - m^2_X) \sin^2(\frac{\theta^*_{\ell}}{2})$
the equations for maximum and
minimum values are:
$Q^2_{max} \approx s - m^2_X$ and $Q^2_{min} \approx 0$.
Solving the equations in (\ref{invar1}) one obtains the
following equations:
\begin{eqnarray}
\theta_{\ell} & = &\, 2\times \arctan \left\{ \pm
\sqrt{\frac{xE_pQ^2}{E_e(xs-Q^2)}}\, \right\} \;,\label{relat1}\\
x\,\, & \geq & \frac{Q^2}{s}\;. \label{relat2}
\end{eqnarray}
The equation (\ref{relat2}) together with
the minimal observable momentum transfer
determines the lowest posible $x$ which can be studied at the
$e^-p^+$-colliders. For example
the smallest $x$ value at the TEVATRON-TESLA PRELC
would be $x \agt 1.1\times 10^{-5}$ ($Q^2 \geq 10\: {\rm GeV}^2$).

\begin{figure}[h]
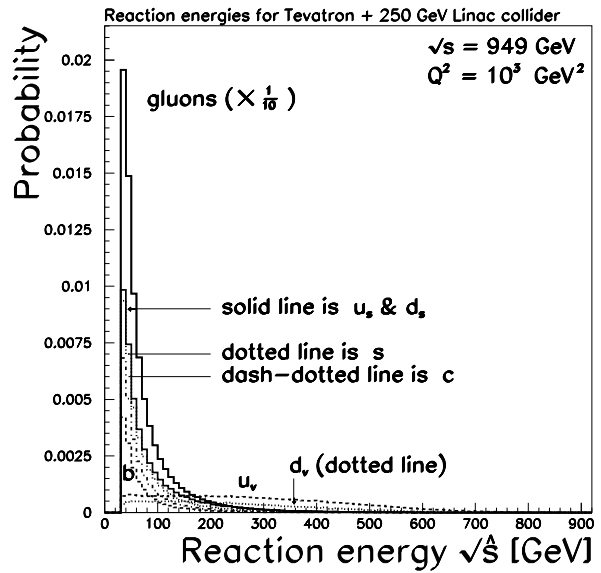

\caption{\rm Probability distributions for different types of reactions
for the momemtum transfer $Q^2=10^3\; {\rm GeV}^2$.}
\label{fig:q103}
\end{figure}

\begin{figure}[h]
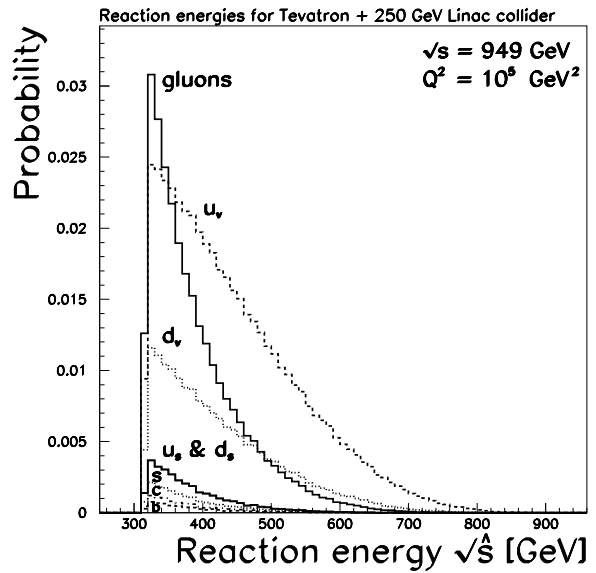

\caption{\rm Same as Fig. \protect\ref{fig:q103} but for
$Q^2=10^5\; {\rm GeV}^2$.}
\label{fig:q105}
\end{figure}

The quark, antiquark and gluon energies $E^i_j$ ($j=q,\bar{q},\;
\text{and}\; g$)
of the proton
are given by $E^i_j = x^i_jE_p$ ($i=1,\ldots$), and
hence the actual reaction energy is $\sqrt{\hat{s}}$,
where $\hat{s} =
 x^i_j s$.
The total reaction energies for different PRELC design
options and  actual
reaction energies for
TEVA\-TRON-TESLA PRELC are shown in Figs. \ref{fig:tots},
\ref{fig:q103} and \ref{fig:q105}. The quark and gluon distributions are
calculated using the PDFLIB program \cite{PDFLIB}.

In the Table  \ref{T:phys_ep}
we have listed
some
interesting physical issues for TeV-range
electron-proton collider (for a more general review see Refs.
\cite{EPPHYS,EPPHYLHC}) and the corresponding ones for the TeV-range
photon-proton collider are presented in Table \ref{T:phys_gp}
(see also Ref. \protect\cite{gammaphys}).
Using the probability distributions shown in Figs.~\ref{fig:q103} and
\ref{fig:q105} one can study the possible reactions (e.g. which reactions are
possible for certain values of $Q^2$, etc.) and some physical
properties of more
specified reactions, for example, the quark contributions to the reaction
$e^- p^+ \rightarrow \nu_e + n$
at very high energies.

The remarkable feature of the PRELC is that the
electron-proton (photon-proton) collisions are not very asymmetric
contrary to
HERA (see Table
\ref{T:pos_option}). This means that the CM-frame is not
moving so fast in the laboratory frame ($1.04 < \gamma_{\rm CM} < 1.67$),
which makes the experimental setup more flexible than it is at
HERA.

In conclusion we have shown that the ring-linac type $e^-p^+$-colliders
will give us a unique and exciting opportunity to study (with reasonably
high luminosities)
TeV-scale physics phenomena together with the future
electron-positron linear accelerators and large hadron colliders.
In addition, there
are many physical processes which can only (or better)
be studied in $e^-p^+$ or
$\gamma p^+$-collisions, such as the proton and photon structure functions at
very small $x$,  and the existence of leptoquarks and excited leptons.
In general the proposed
options for the ring-linac type electron-proton or photon-proton colliders
should be investigated properly and one should study the possibilities to build
these types of PRELC's in the near future.

\acknowledgments
The author expresses his gratitude to the town of Rovaniemi
and to Rovaniemen Maalaiskunta
for grants.

\newpage

\begin{table}[t]
\vspace{-0.26cm}
\begin{tabular}{||l|r|r|l||}
Option           &$E_p$ [GeV]
&$\sqrt{s}$ [GeV]&$\gamma_{{}_{CM}}$\\
\hline
HERA(p)+HERA(e) ($30$)&$820.0$
&$313.7$&$2.70970$\\
SPS+LEP$50$      &$450.0$
&$300.0$&$1.66667$\\
SPS+LEP$90$     &$450.0$
&$402.5$&$1.34164$\\
SPS+LINAC$250$   &$450.0$
&$670.8$&$1.04350$\\
HERA(p)+LINAC100&$820.0$
&$572.7$&$1.60639$\\
HERA(p)+LINAC$250$&$820.0$
&$905.5$&$1.18162$\\
TEVATRON+LINAC$100$&$900.0$
&$600.0$&$1.66667$\\
TEVATRON+LINAC$250$&$900.0$
&$948.7$&$1.21221$\\
DI-TEVATRON+LINAC$100$&$1800.0$
&$848.5$&$2.23917$\\
DI-TEVATRON+LINAC$250$&$1800.0$
&$1341.6$&$1.52798$
\end{tabular}
\vspace{0.07cm}
\caption{Possible options for proton-ring and electron(-linac) collider.
}
\label{T:pos_option}
\end{table}

\begin{table}[h]
\squeezetable
\vspace{-0.26cm}
\begin{tabular}{||l|l|l|l|l||}
$N_p/N_e\, 10^{10}$& $f_p/f_e/f\, {\rm kHz}$&
$\sigma^p_x/\sigma^e_x/\sigma_x\,{\rm\mu m}$&
$\sigma^p_y/\sigma^e_y/\sigma_y\,{\rm\mu m}$&
${\cal L}\:{\rm (s\cdot cm)^{-1}}$\\  \hline
$15/5.15$ (a) & $286/8/8$ &
$36/1.0/36$ & $36/0.06/36$
&$76.\times 10^{28}$  \\
$10/5.15$ (b) & $10417/8/8$
& $265/1.0/265$ & $84/0.06/84$
&$3.0\times 10^{28}$
\end{tabular}
\vspace{0.07cm}
\caption{The most important parameters for (a) TEVATRON-TESLA and
(b) HERA-TESLA
proton-ring and
electron-linac colliders. The proton ring parameters are taken from
\protect\cite{PPARA} and the electron linac parameters from
\protect\cite{EPARA}.}
\label{T:posi_param}
\end{table}

\begin{table}[h]
\narrowtext
\vspace{-0.26cm}
\begin{tabular}{||lcl||}
{\bf Standard Model} & & \\
\hline
EW interactions &:& Higgs boson,
                           properties of $W^\pm$ and \\
& & $Z^0$ and electroweak parameters.\\
Strong interactions &:& parton distribution of proton,\\
                    & &  structure functions and
                      running $\alpha_s$.\\ \hline
{\bf Beyond SM} & &\\
\hline
Extended models &:& New currents and gauge bosons.\\
Supersymmetry &:& First generation of superpartners\\
              & &  and gaugino mixing.\\
Compositeness &:& excited fermions and bosons,\\
& & leptoquarks and leptogluons.
\end{tabular}
\vspace{0.07cm}
\caption{The most important physical issues for TeV-range electron-proton
collider.}
\label{T:phys_ep}
\end{table}

\begin{table}[h]
\vspace{-0.26cm}
\begin{tabular}{||lcl||}
{\bf Standard Model} & &\\
\hline
Real photoproduction of &:& elementary particles, heavy\\
&& quarks,
$W^\pm$ and $Z^0$.\\
EW interactions &:& Total cross-sections of
$\gamma p^+$ \\
&& interactions.\\
Strong interactions &:& parton distributions of proton\\
&& and photon.\\ \hline
{\bf Beyond SM} & &\\
\hline
Real photoproduction of &:& New quarks
($4$th generation and\\
&& weak isosinglets).\\
Real photoproduction of &:& New gauge bosons and SUSY\\
&& particles.\\
Real photoproduction of &:& Leptoquarks, excited fermions\\
&& and bosons.
\end{tabular}
\vspace{0.07cm}
\caption{The same as Table \protect\ref{T:phys_ep} but for
photon-proton collider.}
\label{T:phys_gp}
\end{table}

\newpage

\end{document}